\documentclass[12pt,pra,aps,amssymb,amsfonts,amsmath,tightenlines]{revtex4}

\usepackage{graphicx}
\usepackage{amssymb,amsfonts, bbm}
\usepackage{dsfont}
\usepackage{amsmath}

\usepackage{relsize}

\newtheorem{Proposition}{Proposition}

\newcommand{\half}{\mbox{$\textstyle \frac{1}{2}$}}
\newcommand{\ri}{\mathrm{i}}
\newcommand{\re}{\mbox{$\rm e$}}
\newcommand{\rd}{\mbox{$\rm d$}}
\begin{document}

\title{Determination of the L\'evy Exponent\\ in Asset Pricing Models}
\author{George Bouzianis and Lane P. Hughston}

\affiliation{Department of Computing, Goldsmiths College, University of London, New Cross, London SE14 6NW, United Kingdom
}


\begin{abstract}
\noindent 
We consider the problem of determining the L\'evy exponent in a L\'evy model for asset prices given the price data of derivatives. The model, formulated under the real-world measure $\mathbb P$, consists of a pricing kernel $\{\pi_t\}_{t\geq0}$ together with one or more non-dividend-paying risky assets driven by the same L\'evy process.  If $\{S_t\}_{t\geq0}$ denotes the price process of such an asset then $\{\pi_t S_t\}_{t\geq0}$ is a $\mathbb P$-martingale. The L\'evy process 
$\{ \xi_t \}_{t\geq0}$ is assumed to have exponential moments, implying the existence of a L\'evy exponent 
$\psi(\alpha) = t^{-1}\log \mathbb E(\rm e^{\alpha \xi_t})$ for $\alpha$ in an interval $A \subset \mathbb R$ containing the origin as a proper subset.    We show that if the prices of power-payoff derivatives, for which the payoff is $H_T = (\zeta_T)^q$ for some time $T>0$, are given at time $0$ for a range of values of $q$, where $\{\zeta_t\}_{t\geq0}$ is the so-called benchmark portfolio defined by $\zeta_t = 1/\pi_t$, then the L\'evy exponent is determined up to an irrelevant linear term. In such a setting, derivative prices embody complete information about price jumps: in particular, the spectrum of the price jumps can be worked out from current market prices of derivatives. More generally, if $H_T = (S_T)^q$ for a general non-dividend-paying risky asset driven by a L\'evy process, and if we know that the pricing kernel is driven by the same L\'evy process, up to a factor of proportionality, then from the current prices of power-payoff derivatives we can infer the structure of the L\'evy exponent up to a transformation $\psi(\alpha) \rightarrow \psi(\alpha + \mu) - \psi(\mu) + c \alpha$, where $c$ and $\mu$ are constants. 
\vspace{-0.2cm}
\\
\begin{center}
{\scriptsize {\bf Keywords: Asset pricing; L\'evy models; L\'evy processes; L\'evy exponent; exponential moments;\\ 
option pricing; option replication; power payoffs.
} }
\end{center}
\end{abstract}

\maketitle

\section{Introduction}
\label{Introduction} 

\noindent We are concerned with determining the extent to which derivative prices, taken at time zero, can be used to infer the nature of the underlying jump processes in models where asset prices can move discontinuously. To this end we consider the case of geometric L\'evy models,  and address the question of to what degree the present values of derivatives can be used to determine the L\'evy processes driving the prices of the underlying financial assets. Since the work of Breeden \& Litzenberger (1978) and Dupire (1994), a burgeoning literature has developed based on the idea that given the prices of options and other derivatives one can infer distributional and dynamical properties of the price processes of the underlyings. Most of this work deals with continuous price processes. In the present work we consider discontinuous processes and show that in the case of exponential L\'evy models the L\'evy exponent can be completely determined modulo a two-parameter family of transformations. The paper is structured as follows. In Section II we summarize a few facts concerning L\'evy processes. Then we introduce the condition that the L\'evy process should admit exponential moments and explore some of the implications of this assumption. In Section III we introduce the class of geometric L\'evy models. These models generalize the standard geometric Brownian motion model. They enable one to see the form that the excess rate of return takes as a function of the level of risk, measured by a volatility parameter $\sigma$, and the level of market risk aversion, measured by a parameter $\lambda$. In Section IV we state the framework we use for pricing derivatives and indicate how dividends are handled. In Section V we present a method for determining the underlying jump process given a family of power-payoff derivative prices, and the result is summarized in Proposition \ref{determination of Levy exponent}. Then in Section VI we show in Proposition \ref{Levy exponent via natural numeraire} that if the asset on which the derivatives are based is known to be the natural numeraire, then the underlying jump process can be determined with greater precision. We elaborate further on  the representation of the L\'evy exponent in terms of the prices of power-payoff derivatives and  comment  in particular on the feasibility of using call option prices for a similar purpose. Finally, in Section VII we establish  analogous results for imaginary power-payoff derivatives, where we make use of the techniques of Fourier analysis to show that a general European-style derivative can be expressed as a portfolio of imaginary power payoff derivatives, providing  the payoff is smooth and has good asymptotic properties.

\section{Exponential Moments}
\noindent We shall assume that the reader is 
familiar with L\'evy processes and their financial applications (Andersen \& Lipton 2013, Appelbaum 2004, Bertoin 2004, Brody, Hughston \& Mackie 2012, Chan 1999, Cont \& Tankov 2004, Gerber \& Shiu 1994, 
Hubalek \& Sgarra 2006, Kyprianou 2006, Norberg 2004, Protter 2005, Sato 1999, Schoutens 2004). 
We mainly work with one-dimensional L\'evy processes. For convenience we recall some definitions and classical results. A random process  $\{\xi_t\}$ taking values in $\mathbb R$ on a probability space $(\Omega,{\mathcal F},{\mathbb P})$ is said to be a  L\'evy process if  
(a) $\xi_{s+t}-\xi_s$ and $\{ \xi_u \}_{0 \leq u \leq s}$ are independent for $s,t \geq 0$ (independent increments), 
(b) $\xi_{s+t} - \xi_{s}$ has the same law as $\xi_{t}$ for $s,t \geq 0$ (stationary increments), 
(c) $\lim_{t \rightarrow 0} \mathbb{P}( |\,\xi_{s+t} - \xi_s\,| > \epsilon)=0$ for $\epsilon > 0$ (continuity in probability), and
(d) there exists an $\Omega' \in   {\mathcal F}$ satisfying $ {\mathbb P}(\Omega') =1$ such that for 
$\omega \in \Omega'$ the path $\{ \xi_t(\omega) \}_{t\geq 0} $ is right-continuous for $t\geq 0$ and has left limits for $t>0$ (c\`adl\`ag property). Note that (b) implies that $\xi_0 = 0$ almost surely. It follows from this definition that for $t \geq 0$ and $\kappa \in \mathbb R$ the Fourier transform of $\xi_t$ can be represented in the form
\begin{eqnarray}
\frac{1}{t} \log {\mathbb E} \left[ \exp (\ri \kappa \xi_t)\right] 
= \ri p\kappa - \frac{1}{2}q\kappa^2+ \int_{-\infty}^\infty(\re^{ \ri \kappa x}-1- \ri \kappa x{\mathds 1}\{|x|<1\})\, \nu(\rd x). 
\label{L-K Imaginary}
\end{eqnarray}
Here $p$ and 
$q>0$ are constants and $\nu(\rd x)$ is a L\'evy measure. A Borel measure $\nu(\rd x)$ on $\mathbb R$ is called a L\'evy measure if $\nu(\{0\}) = 0$ and 
\begin{eqnarray}
\int_{-\infty}^\infty 1\wedge x^2\, \nu(\rd x) < \infty, 
\end{eqnarray}
where $a \wedge b = \min (a,b)$. The L\'evy measure associated with a L\'evy process has the property that for any measurable set $B \subset {\mathbb R}$ the expected rate at which jumps occur for which the jump size lies in $B$ is $\nu(B)$. The sample paths of a L\'evy  process have bounded variation on every compact interval of time almost surely if and only if $q = 0$ and 
\begin{eqnarray}
\int_{-\infty}^\infty  \, 1 \wedge  |x| \, \nu(\rd x) < \infty. 
\label{bounded variation}
\end{eqnarray}
In that case we say that $\{\xi_t\}$ has bounded variation. Let us write $\xi_{t-} = \lim_{s \to t} \xi_s$ for the left limit of the process at time $t$. The discontinuity at time $t$ is then defined by $\Delta \xi_t = \xi_t - \xi_{t-} $, and for the L\'evy measure we have
\begin{eqnarray}
\nu(B) = \frac{1}{t} \, \mathbb E  \sum_{0 \leq s \leq t} \mathds 1 \{\Delta \xi_s \in B\}  
\label{bounded variation 2}
\end{eqnarray}
for any $t>0$. If $\nu(\mathbb R) < \infty$ we say that $\{\xi_t\}$ has finite activity, whereas if $\nu(\mathbb R) = \infty$ we say that $\{\xi_t\}$ has infinite activity. A necessary and sufficient condition for (\ref{bounded variation}) to hold is that
\begin{eqnarray}
\sum_{0 \leq s \leq t} | \Delta \xi_s |< \infty 
\label{bounded variation 3}
\end{eqnarray}
almost surely for every $t>0$. If $\sup_t | \Delta \xi_t | \leq c$ almost surely for some constant $c>0$, then we say that $\{\xi_t\}$ has bounded jumps. 

In order for $\{\xi_t\}$ to give rise to a L\'evy model for asset prices, we require additionally that for every $t > 0$ the random variable $\xi_t$ should satisfy a moment condition of the form
\begin{eqnarray}
{\mathbb E}( \re^{\alpha \xi_t} ) < \infty 
\label{moment condition}
\end{eqnarray}
for  $\alpha$ in an interval $A = (\beta, \gamma) \subset \mathbb R$ 
containing the origin. Here we set $\beta = \inf \alpha : {\mathbb E}( \re^{\alpha \xi_t} ) < \infty$ and
$\gamma = \sup \alpha : {\mathbb E}( \re^{\alpha \xi_t} ) < \infty$.
If a L\'evy process satisfies this condition, we say it possesses exponential moments. In that case, 
it follows (Sato 1999, theorem 25.17) that there exists  a so-called L\'evy exponent  
$\psi : \mathbb C_A \to \mathbb R$, 
for 
$ \mathbb C_A = \{\alpha \in \mathbb C \, | \, {\rm Re}(\alpha) \in A\}$,  
such that 
\begin{eqnarray}
{\mathbb E}(\re^{\alpha \xi_t}) = \re^{\psi(\alpha) t},  
\label{cumulant}
\end{eqnarray}
where $\psi(\alpha)$ admits a L\'evy-Khinchin representation of the form
\begin{eqnarray}
\psi(\alpha)=p\alpha+\frac{1}{2}q\alpha^2+
\int_{-\infty}^\infty(\re^{\alpha x}-1-\alpha x{\mathds 1}\{|x|<1\})\, \nu(\rd x). 
\label{L-K}
\end{eqnarray}
A necessary and sufficient condition for a L\'evy process to satisfy (\ref{moment condition}) for $\alpha \in A$ is that 
the associated L\'evy measure should satisfy 
\begin{eqnarray}
\int_{-\infty}^\infty \re^{\alpha x} \, {\mathds 1}\{|x|>1\} \, \nu(\rd x) < \infty 
\end{eqnarray}
for $\alpha \in A$ (Sato 1999, theorem 25.3). If $\{\xi_t\}$ admits exponential moments, then one can check that for $p >0$ we have
\begin{eqnarray}
{\mathbb E}(\, | \xi_t|^p)  < \infty.
\end{eqnarray}
The argument is as follows. Now, for any $\alpha \in \mathbb R$ we have
\begin{eqnarray}
 \cosh(\alpha \xi_t) = \sum_{k = 0}^{\infty} \frac{\, \, \, (\alpha \xi_t)^{2k}} {(2k)!}. 
\end{eqnarray}
Therefore for any $k \in \mathbb N$ we have
\begin{eqnarray}
 \cosh(\alpha \xi_t) >  \frac{\,\, \,(\alpha \xi_t)^{2k}} {(2k)!}. 
\label{moments}
\end{eqnarray}
If we choose $\alpha$ so that $|\alpha| < \min ( |\beta| , \gamma) $, which ensures that $\alpha$ and $-\alpha$ are in $A$, then ${\mathbb E}(  \cosh(\alpha \xi_t) ) < \infty$. Therefore,
${\mathbb E}(\,  |\xi_t|^n)  < \infty$ for even $n \in \mathbb N$, which implies that 
${\mathbb E}(\,  |\xi_t|^p)  < \infty$ for all $p \in \mathbb R^+$, since  for each $n$ and any random variable $X$ it holds that
${\mathbb E}(\,  |X|^{n})  < \infty$ implies ${\mathbb E}(\, |X|^p)  < \infty$ for $0\leq p \leq n$.
More generally, for any $\alpha \in A$ and any $p \in \mathbb R^+$ we have 
\begin{eqnarray}
{\mathbb E}(\,  \re^{\alpha \xi_t} \, | \xi_t|^p )  < \infty.
\label{mixed moments}
\end{eqnarray}
This can be seen as follows. Since $A$ is open, for any $\alpha \in A$ we can choose $\epsilon >0$ so that
$\alpha (1 + \epsilon)$ is still in $A$. Then by H\"older's inequality we have 
\begin{eqnarray}
{\mathbb E}(\,  \re^{\alpha \xi_t} \, | \xi_t|^p )  
\leq  \left( {\mathbb E}(\,  \re^{\alpha (1 + \epsilon) \xi_t} ) \right)^{1/(1 + \epsilon)}  
\left(  {\mathbb E}(\,  | \xi_t|^{p (1 + \epsilon)/ \epsilon )} ) \right)^{\epsilon/(1 + \epsilon)} .
\end{eqnarray}
But we have already established that the terms on the right are finite, and that gives (\ref{mixed moments}). A similar argument shows that if $\{\xi_t\}$ admits exponential moments then  
\begin{eqnarray}
\int_{-\infty}^\infty \re^{\alpha x} \,  |x|^p \, {\mathds 1}\{|x|>1\} \, \nu(\rd x) < \infty, 
\label{mixed Levy moments}
\end{eqnarray}
for $\alpha \in A$ and $p>0$. Setting $\alpha = 0$ and $p = 1$, we see in particular that 
\begin{eqnarray}
\int_{-\infty}^\infty  |x| \, {\mathds 1}\{|x|>1\} \, \nu(\rd x) < \infty, 
\end{eqnarray}
which implies that one can extend the truncated term on the right side of
(\ref{L-K}) to an integral of the form $\int |x| \, \nu(\rd x) $, over the whole of 
$\mathbb R$, by dropping the indicator function and redefining the constant $p$ in equation
(\ref{L-K}). The finiteness of integrals (\ref{mixed moments}) and (\ref{mixed Levy moments}) allows one to compute the Greeks for various derivative payouts in exponential L\'evy models.

Examples of L\'evy processes admitting exponential moments include: (a) Brownian motion, for which $\psi(\alpha) = \half \alpha^2$, $\alpha \in \mathbb R$; (b) the Poisson process with rate $m$, for which $\psi(\alpha) = m(\re^{\alpha} - 1)$ and $\nu(\rd z) = m \delta_{1}(\rd z)$; (c) the compound Poisson process with rate $m$, for which  $\psi(\alpha) = m(\theta(\alpha)- 1)$, where $\theta(\alpha)$ is the moment generating function for the distribution 
$\mu(\rd x)$ of a typical jump and $\nu(\rd z) = m \mu(\rd x)$; (d) the gamma process with rate $m$, for which $\psi(\alpha) = -m\log (1-\alpha)$, $\alpha < 1$, where $\nu(\rd z) = \mathds 1\{ {z>0} \} \, z^{-1} \exp( -z) \, \rd z$ (Dickson \& Waters 1993, Heston 1993, Brody, Macrina \& Hughston 2008, Yor 2007); (e) the variance gamma (VG) process, for which 
$\psi(\alpha) = -m\log (1-\alpha^2 / 2 m^2)$, where we have $-2^{1/2} m < \alpha < 2^{1/2} m$ (Madan \& Seneta 1990, Madan \& Milne 1991, Madan, Carr \& Chang 1998); (f) the truncated stable family of L\'evy processes, which includes the gamma process and the VG process as special cases (Koponen 1995, Carr, Geman, Madan \& Yor 2002, Andersen \& Lipton 2013, K\"uchler \& Tappe 2014); (g) hyperbolic processes (Eberlein \& Keller 1995, Eberlein, Keller \& Prause 1998, Bingham \& Kiesel 2001); (h) generalized hyperbolic processes (Eberlein 2001); (i) normal inverse Gaussian processes (Barndorff-Nielsen 1998); and (j) Meixner processes (Schoutens \& Teugels 1998).

\section{Geometric L\'evy Model}
\label{GLMl} 

\noindent The geometric L\'evy model for asset prices can be viewed as an extension of the well-known geometric Brownian motion model to the L\'evy regime.  For simplicity, we consider a model driven by a one-dimensional process $\{ \xi_t \}_{t \geq 0}$. The generalization to higher dimensional L\'evy processes is straightforward. We assume that $\{ \xi_t \}$ admits exponential moments and denote the associated L\'evy exponent by $\{\psi(\alpha)\}_{\alpha \in A}$ for $A = (\beta, \gamma)$ with $\beta < 0 < \gamma$. The process $\{m_t\}_{t \geq 0}$
defined by
\begin{eqnarray}
m_t = \re^{\alpha \xi_t- \psi(\alpha) t },   
\label {martingale}
\end{eqnarray}
for some choice of $\alpha \in A$, is the corresponding
geometric L\'evy martingale with volatility $\alpha$. 
By the stationary and independent increments properties we
find that ${\mathbb E}_s (m_t) = m_s$.
Here we write $\mathbb E_t(\cdot) = \mathbb E \, (\cdot \, | \, \mathcal F_t)$, where  $\{{\mathcal F}_t\}$ denotes the augmented filtration generated by $\{\xi_t\}$.
The associated geometric L\'evy model consists of a pricing kernel, a money market account, and one or more so-called investment-grade assets. See Duffie (1992), Hunt \& Kennedy (2004), Cochrane (2005) for general aspects of the theory of pricing kernels in arbitrage-free asset pricing models. For the construction of the pricing kernel $\{\pi_t\}_{t\geq0}$ in the context of a L\'evy model we let $r \in \mathbb R$ and $\lambda>0$ be constants, and assume that 
$-\lambda\in A$. Then we set 
\begin{eqnarray}
\pi_t = \re^{ -rt -\lambda \xi_t - \psi(-\lambda) t}.
\label{levy pricing kernel}
\end{eqnarray}
We refer to the related process $\{ \zeta_t\}_{t \geq 0}$ defined by $\zeta_t = 1/\pi_t$ as the growth-optimal portfolio or {\em natural numeraire} asset (Flesaker \& Hughston 1997). It serves as a benchmark relative to which other non-dividend-paying assets are martingales. In some calculations it is convenient to make reference to the natural numeraire instead of the pricing kernel. The money market account $\{B_t\}_{t\geq0}$ is taken to have the value $B_t = B_0 \re^{rt}$ at time $t$, where $B_0$ denotes its initial value in some fixed unit of account.  The idea of an investment-grade asset is that it should offer a rate of return that is strictly greater than the interest rate. Ordinary stocks and bonds are in this sense investment-grade, whereas put options and short positions in ordinary stocks and bonds are not. We assume for the moment that the assets pay no dividends over the time horizons considered (dividends will be treated shortly), and we write $\{S_t\}_{t\geq0}$ for the value process of a typical non-dividend paying risky asset in the geometric 
L\'evy model. We require that the product of the pricing kernel and 
the asset price should be a martingale, which we take to be geometric L\'evy martingale of the form
\begin{eqnarray}
\pi_tS_t = S_0\re^{\beta \xi_t - \psi(\beta) t}
\end{eqnarray}
for some $\beta\in A$. From the formulae above we deduce that
\begin{eqnarray}
S_t 
= S_0 \, \re^{rt + \sigma \xi_t+\psi(-\lambda) t-\psi(\sigma-\lambda) t},
\label {price}
\end{eqnarray}
where $\sigma = \beta + \lambda$. We assume that  $\sigma > 0$ and that 
$\sigma\in A$. It follows that the price can be expressed 
in the form
\begin{eqnarray}
S_t = S_0 \, \re^{rt + R(\lambda,\sigma)t + \sigma \xi_t-\psi(\sigma) t},
\label{asset price}
\end{eqnarray}
where
\begin{eqnarray}
R(\lambda,\sigma) = \psi(\sigma) + \psi(-\lambda) - \psi(\sigma-\lambda). 
\label{R} 
\end{eqnarray}
Thus we see that $\sigma$ is the volatility of the asset price relative to the given L\'evy process and that $R(\lambda,\sigma)$ is the excess rate of return above the interest rate. The parameter $\lambda$ can be interpreted as a measure of the level of market risk aversion. A calculation shows that the excess rate of return is bilinear in $\lambda$ and $\sigma$ if and only if $\{\xi_t\}$ 
is a Brownian motion (Brody \textit{et al} 2012). It follows that the interpretation of $\lambda$ as a ``market price of risk'', which is valid for models based on a Brownian filtration, does not 
carry through directly to the general L\'evy regime. Nevertheless, the notion of excess rate of return is well defined, 
and under the assumptions that we have made the strict convexity of the L\'evy exponent implies that  the excess rate of return is strictly positive. To show that $R(\lambda,\sigma) >0$ we can use the 
L\'evy-Khinchin formula (\ref{L-K}) to deduce that 
\begin{eqnarray}
R(\lambda,\sigma) = 
\int_{-\infty}^\infty (\re^{\sigma x}-1)(1 - \re^{-\lambda x}) \, \nu(\rd x). 
\label{excess}
\end{eqnarray}
It follows by inspection of  (\ref{excess}) that the excess rate of return is an increasing function of the volatility and the level of risk aversion. 

In the case of a single asset driven by a single L\'evy process one can without loss of generality set $\sigma = 1$. This can be achieved by defining a rescaled L\'evy process $\{\bar \xi_t\}$ by setting $\bar \xi_t = \sigma \xi_t$. Then we define $\bar \psi (\alpha) = \psi(\sigma \alpha)$ and set $\bar \lambda = \lambda / \sigma$, and we have 
\begin{eqnarray}
\pi_t = \re^{ -rt -\bar \lambda \bar \xi_t - \bar \psi(-\bar \lambda) t}, \quad S_t = S_0 \, \re^{rt + \bar R(\bar \lambda,1)t + \bar \xi_t- \bar \psi(1) t},
\label{levy pricing kernel rescaled}
\end{eqnarray}
where $\bar R(\bar{\lambda},1) = \bar \psi(1) + \psi(- \bar \lambda) - \bar \psi(1- \bar \lambda)$. If we then drop the bars, we are led back to a version of the model already set up, but with $\sigma = 1$. Nevertheless, it can be helpful to leave the parameter $\sigma$ intact as part of the theory, since there are situations where one would like to compare versions of the model for different values of the parameter, e.g. for sensitivity analysis and calculations of the Greeks. On the other hand, there are also situations where it is desirable to make use of simplifications resulting from setting $\sigma = 1$; an example of this can be found in the proof of Proposition  \ref{determination of Levy exponent}.
It should be noted that the value of the asset given by \eqref{asset price} does not depend on the drift of the L\'evy process, for if we replace $\xi_t$ with $\xi_t + \epsilon t$ for some $\epsilon \in \mathbb R$ then the L\'evy exponent $\psi(\alpha)$ gets replaced with $\psi(\alpha) + \epsilon \alpha$, and the combination $\sigma \xi_t - \psi(\sigma) t$ appearing in the formula for the asset price is left unchanged. This has the implication that if one attempts to determine the L\'evy exponent from the prices of derivatives, one will be left with an indeterminacy of the form $\psi(\alpha) \to \psi(\alpha) + c\alpha$ for some unknown constant $c$. 

It is worth recalling that one of the motivations indicated by Mandelbrot (1963)  for the introduction of L\'evy models in finance is the possibility of offering an explanation for the apparent existence of ``fat tails'' in the distributions of returns. But it seems that what he had in mind was not the construction of specific dynamical models for price processes, but rather the introduction of infinitely-divisible distributions with infinite moments to model the returns on such assets, an assumption that makes the construction of dynamical models difficult.  From an empirical point of view, however, the requirement a L\'evy process should have ``thin tails'' is a relatively mild one: a sufficient condition for a L\'evy process to admit exponential moments (and hence to have thin tails) is that the jumps should be bounded (Protter 2005, theorem 34), even if the bounds are set at arbitrarily high values. Thus, from a modern point of view the use of L\'evy processes in finance stems not so much from a desire to model the distributions of the tails of returns but rather to account for the characteristics of the jumps that asset prices can undertake.

\section{Derivative pricing}
\noindent The price $H_0$ at time $0$ of a European style derivative that delivers a single random payment $H_T$ at time 
$T$ is given by
\begin{eqnarray}
H_0 = \mathbb E (\pi_T H_T).
\label{derivative price} 
\end{eqnarray}
The expectation is, of course, with respect to the real-world probability measure. The pricing kernel takes care of both the discounting and the probability weighting needed to give the answer. The use of such a formula for derivative pricing is well known, but it may be useful to recall the argument. In the general theory of asset pricing one fixes, as we have done, a probability space $(\Omega, \mathcal F, \mathbb P)$, where $\mathbb P$ is interpreted as the real-world measure, together with a filtration $\{ \mathcal F_t \}$, and one assumes the existence of a pricing kernel $\{\pi_t\}_{t\geq 0}$ satisfying $\pi_t >0$ for all $t\geq0$ such that for any asset with a non-negative price process $\{S_t\}_{t\geq0}$ and a non-decreasing cumulative dividend process $\{K_t\}_{t\geq0}$, the associated deflated gain process
$\{\bar{S}_t\}_{t\geq0}$ defined by
\begin{equation} \label{deflated total value process}
	\bar{S}_t = \pi_t S_t + \int_0^t \pi_s \,\rd K_s
\end{equation}
is a martingale. So far, we have considered limited liability assets, for which the prices are non-negative and the cumulative dividend process is increasing. By a general asset, not necessarily of limited liability, we mean an asset with the property that its price can be expressed as the difference between the prices of two limited liability assets, and for which the  cumulative dividend process can be expressed as the difference between two increasing cumulative dividend processes. It follows then that the deflated gain process of a general asset is also a martingale. 

The formula given above allows for the possibility of both continuously paid and discretely paid dividends. If the dividends are entirely discrete (and paid at possibly random times), then the deflated gain process can be expressed in the form
\begin{equation} \label{discrete deflated total value process}
	\bar{S}_t = \pi_t S_t + \sum_{0 \leq s \leq t} \pi_s \, \Delta(K_s).
\end{equation}
At each time at which a dividend is paid, the cumulative dividend process jumps, and the value of the jump is equal to the dividend. In the case of a European style derivative with a single payoff $H_T$ made at time $T$, we think of the payoff as a dividend, and hence the cumulative dividend process is zero up to time $T$, then jumps to $H_T$ at $T$. The sum in (\ref{discrete deflated total value process}) reduces to a single term, given by the jump $\Delta(K_T) = H_T$, and we have $\bar{S}_T= \pi_T H_T$ at $T$. Since the value of the derivative itself drops to zero the instant that the dividend is paid, it follows by the martingale condition that 
$\bar{S}_0 = \mathbb E (\pi_T H_T)$, which gives us (\ref{derivative price}). In the literature, by virtue of a conventional abuse of notation, one often writes the price process of the derivative in the form $\{H_s\}_{0 \leq s \leq T}$, as if somehow the terminal value of the derivative is what is paid; in reality, the value of the derivative itself at maturity is 0, whereas the payoff (or dividend) at time $T$ is $H_T$. This is consistent with the notion that the price process of the derivative should be right continuous with left limits.

We are also in a position to check that if a risky asset pays a proportional dividend at the constant rate $\delta$ then its price in the geometric L\'evy model will be given by 
\begin{eqnarray}
S_t = S_0 \, \re^{(r-\delta) t + R(\lambda,\sigma)t + \sigma \xi_t-\psi(\sigma) t}.
\label{asset price with dividends}
\end{eqnarray}
The expression is of course intuitively plausible, perhaps even obvious, by analogy with the corresponding result in the geometric Brownian motion model. Nevertheless, we need to check that the process $\{ \bar{S}_t \}$ defined by (\ref{deflated total value process}) is a martingale. A calculation making use of equations
(\ref{levy pricing kernel}), (\ref{deflated total value process})  and 
(\ref{asset price with dividends}) gives
\begin{eqnarray} 
\bar{S}_t  = S_0 \, \re^{-\delta t + (\sigma - \lambda)\xi_t  -\psi(\sigma - \lambda)t} + \delta S_0 \int_0^t \re^{-\delta u + (\sigma - \lambda)\xi_u - \psi(\sigma - \lambda)u} \rd u.
\end{eqnarray}
Splitting the integral at time $s < t$ and taking a conditional expectation making use of Fubini's theorem, we get
\begin{align} 
& \mathbb E_s \bar{S}_t  = S_0 \, \re^{-\delta t + (\sigma - \lambda)\xi_s  -\psi(\sigma - \lambda)s} 
+ \delta S_0 \int_0^s \re^{-\delta u + (\sigma - \lambda)\xi_u - \psi(\sigma - \lambda)u} \rd u 
\nonumber\\ 
&\hspace{2cm} +  \delta S_0 \int_s^t \re^{-\delta u 
 + (\sigma - \lambda)\xi_s - \psi(\sigma - \lambda)s} \rd u, 
\end{align}
from which it follows immediately that $\mathbb E_s \bar{S}_t = \bar{S}_s$. Thus we have shown that the price defined by (\ref{asset price with dividends}) together with the proportional dividend rate $\delta$ is such that the resulting deflated gain process is a martingale, which demonstrates that (\ref{asset price with dividends}) is indeed the correct expression in the geometric L\'evy model for the price of a risky asset that pays a proportional dividend at a constant rate. 

\section{Determination of the L\'evy exponent}

\noindent Turning to the problem of the determination of the L\'evy exponent from price data, we proceed to consider a one-parameter family of so-called power-payoff derivatives, for which 
\begin{eqnarray}
H_T = (S_T)^q, 
\end{eqnarray}
where $q \in \mathbb R$. Various authors have considered the pricing of similar structures in L\'evy models relating the L\'evy exponent to the price (see, e.g., Carr \& Lee 2009, Fajardo 2018, and references cited therein). The value $H_0 \in \mathbb{R^+} \cup \infty$ of a power-payoff derivative at time zero is given by
\begin{eqnarray}
H_0 = \mathbb E (\pi_T S_T^{\,q}).
\label{power price} 
\end{eqnarray}
Here we allow the possibility that the value of the derivative may not be finite for some values of $q$. In a model driven by Brownian motion, the asset price takes the form
\begin{eqnarray}
S_T 
= S_0 \, \re^{(r+\lambda \sigma) T + \sigma W_T - \frac{1}{2} \sigma^2T},
\label {Brownian price}
\end{eqnarray}
and for the pricing kernel we have
\begin{eqnarray}
\pi_T 
=  \re^{-rT - \lambda W_T - \frac{1}{2} \lambda^2 T},
\label {Brownian pricing kernel}
\end{eqnarray}
where for convenience we set $\pi_0 = 1$. It follows that
\begin{eqnarray}
\pi_T S_T^{\,q} 
= S_0^{\,q} \, \re^{(q-1)rT+q\lambda \sigma T + (q\sigma - \lambda) W_T 
- \frac{1}{2}(q \sigma^2 + \lambda^2) T}.
\label {product}
\end{eqnarray}
A calculation then allows one to deduce that the value of the power payoff derivative, regarded as a function $q$, takes the form
\begin{eqnarray}
H_0(q) = S_0^{\,q} \, \re^{(q-1) rT+\frac{1}{2} q (q-1) \sigma^2 T}
\label{power price q} 
\end{eqnarray}
in the Brownian case. One notes that the terms involving $\lambda$ cancel when the expectation is taken, so the value of the derivative only depends on $S_0$, $r$, $\sigma$, and $q$, and that $H_0(q)$ is finite for all values of $q$. 

In the case of a geometric L\'evy model, the asset price is given by (\ref{asset price}) and the pricing kernel is given by (\ref{levy pricing kernel}). Thus we have
\begin{eqnarray}
\pi_T S_T^{\, q} 
= S_0^{\, q}  \, \re^{(q-1)rT + (q\sigma -\lambda) \xi_T+ (q-1)\psi(-\lambda) T
-q \psi(\sigma-\lambda) T}.
\end{eqnarray}
The value of the power-payoff derivative regarded as a function of $q$ then takes the form
\begin{eqnarray}
H_0 (q)
= S_0^{\, q}  \, \re^{(q-1)rT + \psi(q\sigma -\lambda)T + (q-1)\psi(-\lambda) T
-q \psi(\sigma-\lambda) T}.
\end{eqnarray}
It is perhaps remarkable that an explicit expression is obtained, but this allows one to study in detail the relation between the type of L\'evy model under consideration and the resulting values of derivatives. 
We note that $H_0 (0) = \re^{-rT}$ and that $H_0 (1) = S_0$, as one would expect, and  it should be evident that in general $H_0 (q)$ is finite only for a certain range of values of the parameter $q$. In particular, suppose that $\sigma > \lambda > 0$ and that $\sigma \in A$ and $-\lambda \in A$.  Then for $A = (\beta, \gamma)$ clearly $q\sigma - \lambda \in A$ if and only if 
\begin{eqnarray}
\frac{1}{\sigma} (\beta + \lambda) < q < \frac{1}{\sigma} (\gamma + \lambda). 
\label{q inequality}
\end{eqnarray}
Since $\beta < -\lambda$ and $\gamma > 0$, these inequalities ensure that the interior of the set of values of $q$ for which 
$H_0 (q) < \infty$ is an open set $B$ that includes the origin.

Now we are in a position to ask to what extent specification of the family of derivative prices $\{H_0 (q)\}_{q \in B}$ allows one to infer the nature of the L\'evy process driving the model. To this end we note that from observations of $H_0 (0)$ and $H_0 (1)$ one can infer the value of $r$ and $S_0$. Thus without loss of generality it suffices to regard the function 
\begin{eqnarray}
D_0(q) = \frac {1}{T} \log \frac{H_0 (q)}{S_0^{\, q}  \, \re^{(q-1)rT } } =\psi(q\sigma -\lambda) + (q-1)\psi(-\lambda) -q \psi(\sigma-\lambda) ,
\end{eqnarray}
which is finite for $q \in B$, as representing the data supplied by the family of derivative prices.
%
\begin{Proposition} \label{determination of Levy exponent}
Let the prices of derivatives with payoff payoffs $H_T = (S_T)^q$ for $q \in \mathbb R$ be given for a non-dividend-paying risky asset $\{S_t \}_{t \geq 0}$ that is known to be a geometric L\'evy process, and suppose it is known that the pricing kernel is a geometric L\'evy process driven by the same L\'evy process up to a factor of proportionality. Then the  L\'evy exponent can be inferred up to a transformation $\psi(\alpha) \rightarrow  \psi(\alpha +\mu) - \psi(\mu) + c \alpha$, where $c$ and $\mu$ are constants.
\end{Proposition}

\noindent {\em Proof.} Without loss of generality one can set $\sigma = 1$. Then we have
\begin{eqnarray}
D_0(q) =\psi(q -\lambda) + (q-1)\psi(-\lambda) -q \psi(1-\lambda).
\label{D0}
\end{eqnarray}
In the setting of the problem we take $D_0(q)$ to be given for all $q \in \mathbb R$ and finite in some open set $B$, and we consider $\lambda$ to be unknown. The goal is to determine the L\'evy exponent. Writing $\tilde \psi (\alpha) = \psi (\alpha -\lambda) - \psi (-\lambda)$, we have 
$D_0(q) = \tilde \psi (q) - q \tilde \psi (1)$. This implies that $\tilde \psi (q) = D_0(q) + qb$ for some $b \in \mathbb R$. Now, it is easy to see that  
$\psi (\alpha) = \tilde \psi (\alpha +\lambda) - \tilde \psi (\lambda)$. We conclude that  for some choice of $\lambda$ and $b$ the L\'evy exponent takes the form
\begin{eqnarray}
\psi (\alpha) = D_0 (\alpha +\lambda) - D_0 (\lambda) + b \alpha.
\label{psi in term of D0}
\end{eqnarray}
Substituting (\ref{psi in term of D0}) into the right-hand side of (\ref{D0}), one can check that the solution is valid.  Finally, we note that under a transformation of the form 
$\psi(\alpha) \rightarrow \hat \psi (\alpha)$, 
with $\hat \psi (\alpha) =  \psi(\alpha +\mu) - \psi(\mu) + c \alpha$, 
where $c, \mu \in \mathbb R$, we find that 
$\hat \psi (\alpha) = D_0 (\alpha +\lambda +\mu) - D_0 (\lambda +\mu) + b \alpha$. The effect of the transformation is $\lambda \to \hat \lambda = \lambda + \mu$.   Since $\lambda$ is unknown, this shows that the L\'evy exponent is determined only up to a transformation of the type indicated. 
 \hfill $\Box$
 
\section{Interpretive Remarks}
 
\noindent Following on from this result, a few  comments may be in order. We recall that a L\'evy process is completely characterized by the random variable $\xi_t$ at a single instant of time $t$. This reflects the fact that there is a one-to-one correspondence between L\'evy processes and infinitely divisible distributions, and a L\'evy process has the property that the distribution of its value at any particular time is infinitely divisible. Taking $t=1$ for convenience, we have $\psi(\alpha) = \log \mathbb E [ \re^{\alpha \xi_1} ]$ in the case of a L\'evy process that admits exponential moments, and we note that the distribution of $\xi_1$ is determined by the L\'evy exponent 
 $\{\psi(\alpha)\}_{\alpha \in A}$. Each such distribution belongs in a natural way to a certain one-parameter family of distributions, which we call an Esscher family of distributions. The distribution of $\xi_1$ is the function $F : \mathbb R \to [0, 1]$ defined by  
 $F(x) = \mathbb E [ \mathds 1\{ \xi_1 \leq x\} ]$.
For the associated L\'evy exponent we have 
\begin{eqnarray}
\psi(\alpha)  = \log \int_{-\infty}^{+ \infty}  \re^{ \alpha x}  \,  \rd F(x).
\end{eqnarray}
The corresponding Esscher family  $F_{\delta} : \mathbb R \to [0, 1]$, $\delta \in A$, is defined by the measure change
\begin{eqnarray}
F_{\delta}(x)= \mathbb E [ \rm e^{ \delta\xi_1 - \psi(\delta) } \mathds 1\{ \xi_1 \leq x\} ].
\end{eqnarray}
 We may accordingly ask for the structure of the L\'evy exponent associated with $F_{\delta}$. This is
\begin{eqnarray}
\psi_\delta(\alpha)  = \log  \int_{-\infty}^{+ \infty}  \re^{ \alpha x}  \,  \rd F_{\delta}(x)
= \log  \int_{-\infty}^{+ \infty}  \re^{ \delta x - \psi(\delta) }  \re^{ \alpha x}  \,  \rd F(x) = \psi(\alpha + \delta) - \psi(\delta). 
\end{eqnarray}
So we see that by Proposition \ref{determination of Levy exponent}, the specification of the prices of power-payoff derivatives allows one to determine the Esscher family of the L\'evy exponent of the underlying L\'evy process, modulo an irrelevant linear term.  L\'evy processes that are equivalent in this sense can be said to belong to the same ``noise type" (Brody, Hughston \& Yang 2013). 

On the other hand, if more information is known \textit{a priori} about the nature of the underlying asset, then a more precise determination of the L\'evy exponent is possible. Consider the case, for instance, where it is known that the asset on which the power payoff derivative is based is the natural numeraire. Then $\sigma = \lambda$, and for the asset price we have
\begin{eqnarray}
\zeta_t = \zeta_0 \, \re^{rt + R(\lambda, \lambda)t + \lambda \xi_t-\psi(\lambda) t},
\label{natural numeraire price}
\end{eqnarray}
where the excess rate of return is given by $R(\lambda,\lambda) = \psi(\lambda) + \psi(-\lambda)$. 
In this case we can without loss of generality set $\lambda = 1$. It follows then from (\ref{D0}) and 
(\ref{psi in term of D0}) that $D_0(1) =0$ and hence $\psi (\alpha) = D_0 (\alpha + 1) + b \alpha$.
Thus we have shown the following:
%
\begin{Proposition} \label{Levy exponent via natural numeraire}
Let the prices of derivatives with power payoffs $H_T = (\zeta_T)^q$ for $q \in \mathbb R$ be given for a natural numeraire $\{\zeta_t \}_{t \geq 0}$ that is known to be a geometric L\'evy process. Then the  L\'evy exponent can be inferred up to a transformation of the form $\psi(\alpha) \rightarrow  \psi(\alpha) + b \alpha$, where $b$ is a constant.
\end{Proposition}

In a geometric L\'evy model, the pricing kernel can be written in the form $\pi_t = e^{-r t }\Lambda_t$ where the martingale $\{\Lambda_t\}_{t \geq 0}$ defined by 
\begin{eqnarray}
\Lambda_t = e^{- \lambda \xi_t - \psi(-\lambda) t}
\end{eqnarray}
determines a change of measure. Thus, for any $\mathcal{F}_t$-measurable random variable $Z_t$ we have
\begin{eqnarray}
\tilde{\mathbb{P}}(Z_t < z ) = \tilde{\mathbb{E}}\,[\mathds{1}(Z_t<z)\,] =
 \mathbb{E}\,[\,\Lambda_t\,\mathds{1}(Z_t<z)\,].
\end{eqnarray}
We refer to $\tilde{\mathbb{P}}$ as the risk-neutral measure and write $\tilde{\mathbb{E}}$ for expectation under $\tilde{\mathbb{P}}$. The terminology ``risk-neutral" comes from the fact that $\tilde{\mathbb{E}}(S_t) = S_0 e^{rt}$ in the geometric L\'evy model. Then $\{\tilde{\psi}(a)\}$ has the interpretation of being the L\'evy exponent associated with $\xi_t$ under the risk-neutral measure. That is to say, 
\begin{eqnarray}
\tilde{\psi}(a) = \frac{1}{t}\log \tilde{\mathbb{E}}[e^{a\xi_t}].
\end{eqnarray}
The essence of Proposition \ref{determination of Levy exponent} is that the family of derivative prices can be used to calculate $\{\tilde{\psi}(a)\}$, which fixes the exponent $\{\psi(a)\}$ under $\mathbb{P}$, modulo the freedom indicated.

Let us write $C_{0T}(x)$ for the price at time $0$ of a call option with maturity $T$ and strike $x$. Can one use the data $\{C_{0T}(x)\}_{x \geq 0}$ for fixed $T$ to ascertain the L\'evy exponent in a geometric L\'evy model? The answer is yes, though the method is less straightforward than the use of power-payoff derivatives, as we shall see. Now, it is known that if the random variable $S_T$ corresponding to the terminal value of the asset at time $T$ admits a risk-neutral density function, then we can use the idea of Breeden \& Litzenberger (1978) to work out this density in terms of call option data. In particular, if we write
\begin{eqnarray}
\tilde{\theta}(x) = \frac{\rd}{\rd x} \, \tilde{\mathbb{P}}(S_T \leq x)
\label{Density of S_T under RN}
\end{eqnarray}
for the density of $S_T$ under the risk-neutral measure $\tilde{\mathbb{P}}$, we have 
\begin{eqnarray}
\tilde{\theta}(x) = e^{-r T}\,\frac{\rd^2 C_{0T}(x)}{\rd x^2}.
\label{Density of S_T under RN Breeden Formula}
\end{eqnarray}
This follows from the fact that 
\begin{eqnarray}
C_{0T}(x) = e^{-r T} \int_{0}^{\infty} (y - x)^{+} \, \tilde{\theta}(y) \, \rd y.
\label{Call price with integral}
\end{eqnarray}
Then $\{ \tilde{\theta}(x) \}_{x \geq 0}$ can be used to calculate the values of power-payoff derivatives via the relation
\begin{eqnarray}
\tilde{\mathbb{E}}(S^{\, q}_{T}) = \int_{0}^{\infty}x^q \,\tilde{\theta}(x)\, \rd x ,
\label{Power Payoff Derivative RN}
\end{eqnarray}
and from there we can work out $\tilde{\psi}(\alpha)$, as indicated in the previous section. The difficulty with this approach is that in a geometric L\'evy model the distribution of $S_T$ does not in general admit a density function, and the system of call option prices $\{C_{0\,T}(x)\}_{x \geq 0}$ is not differentiable for all $x \in \mathbb R^+$. The situation can be remedied to some extent if instead we make use of the risk-neutral distribution function $\{\tilde F(x)\}_{x \geq 0}$ and express the option price in the form of a Lebesgue-Stieltjes integral, writing
\begin{eqnarray}
C_{0T}(x) = e^{-r T} \int_{0}^{\infty} (y - x)^{+} \, \rd \tilde F(y),
\label{Call price as Lebesgue-Stieltjes integral}
\end{eqnarray}
with the understanding that the distribution function is right-continuous. Then the derivative of the option price with respect to the strike is defined for all $x \in \mathbb R^+$  apart from points of discontinuity of the distribution function, and this is sufficient to enable us to recover the distribution function in its entirety. Once the distribution function is known, one can determine the L\'evy exponent by calculating the system of power-payoff prices, given for $q \in \mathbb R$ by 
\begin{eqnarray}
H_0(q) = e^{-r T} \int_{0}^{\infty}x^q \,\rd \tilde F(x) .
\label{Power Payoff Derivative Price}
\end{eqnarray}

\section{Imaginary power payoffs}

\noindent As another example of a one-parameter family of derivatives from which information can be extracted concerning the L\'evy exponent when the underlying is a geometric L\'evy asset we consider a family of imaginary power payoffs, for which the terminal cash flow is given by
\begin{eqnarray}
F_T(q)=(S_T)^{iq},
\label{imaginary power payoff}
\end{eqnarray}
where $q \in \mathbb R$. The value of such a contract at time zero takes the form
\begin{eqnarray}
F_0(q) = \mathbb E (\pi_T S_T^{\,iq}) = \mathbb E (\pi_T e^{iq\log S_T}).
\label{Imaginary Derivative}
\end{eqnarray}
Since the payoff is a complex function, we are in effect valuing two different derivatives, each with its own payoff. That is to say,
\begin{eqnarray}
F_0(q) = \mathbb E [ \pi_T  \cos(q\log S_T)] + i\,\mathbb E [ \pi_T  \sin(q\log S_T) ]  .
\end{eqnarray}
Thus $\{F_0 (q)\}$, $q \in \mathbb{R}$, can be thought of as a pair of price families $\{F_0^c (q)\}$ and $\{F_0^s (q)\}$, for which the corresponding payoff functions are given respectively by
$F_T^c (q)=\cos(q\log S_T)$ and $F_T^s (q)=\sin(q\log S_T)$.
Note that the payoffs, and hence the prices, are bounded for all values of $q$.  
A calculation then shows that
\begin{eqnarray}
F_0(q) = S_0^{iq}\, \re^{r(iq-1)T}\, \re^{(-iq \, \psi(\sigma-\lambda) +(iq -1)\psi(-\lambda) + \psi(iq\sigma-\lambda)\, )T }.
\label{value of imaginary power payoff}
\end{eqnarray}

With these ideas at hand we can use the methods of Fourier analysis to investigate more general payoffs. We begin by recalling briefly a few well known facts. Let the map $f:  \mathbb R \to \mathbb R$ be such that $f \in L^1 $. The Fourier transform of $f$ is the function $g:  \mathbb R \to \mathbb C$ defined by
\begin{eqnarray}
g(q) = \frac{1}{\sqrt{2\pi}} \,\int_{-\infty}^{\infty} e^{-iqx}f(x) \, \rd x .
\end{eqnarray}
Under various further conditions the relation between $f$ and $g$  can then be inverted. For example, if $f \in L^1 $ and is continuous, and if $g \in L^1 $, then
\begin{eqnarray}
f(x) = \frac{1}{\sqrt{2\pi}} \, \int_{-\infty}^{\infty}  e^{iqx}g(q) \, \rd q .
\label{Fourier inversion formula}
\end{eqnarray}
A sufficient condition for these requirements to be satisfied is that $f$ should be a ``good" function in the sense of Lighthill (1958), that is to say, that it should be everywhere differentiable any number of times and such that it and all its derivatives are 
$O \left (\, \left | x \right | ^{-n} \right )$ as $\left | x \right | \rightarrow \infty$ for all $n \in \mathbb{N}$. We recall that $f(x)$ is said to be $O(h(x))$ as $x \rightarrow \infty$ if
\begin{eqnarray}
 \limsup _{x \rightarrow \infty} \, \left | \frac{f(x)}{h(x)} \right | < \infty.
\end{eqnarray}
If $f$ is a good function then its Fourier transform $g$ is also good.
Now consider the situation where the payoff of a European-style derivative with value $H_0$ at time zero takes the form $H_T = f(\log S_T)$ for some $f \in L^1$. If $f$ is continuous and $g \in L^1$, then by use of (\ref{Fourier inversion formula}) we can write the payoff in the form 
\begin{eqnarray}
H_T = \frac{1}{\sqrt{2\pi}} \, \int_{-\infty}^{\infty} e^{\,iq\log S_T}g(q) \, \rd q .
\label{Derivative fourier payoffs}
\end{eqnarray}
This expresses $H_T$ as a portfolio of imaginary power payoffs parameterized by $q$, where $g(q)$ determines the relative portfolio weighting for the given $q$. Multiplying each side of equation (\ref{Derivative fourier payoffs}) with the pricing kernel $\pi_T$ and taking the expectation we obtain
\begin{eqnarray}
\mathbb{E}(\pi_TH_T) = \frac{1}{\sqrt{2\pi}} \,\mathbb{E} \left(  \int_{-\infty}^{\infty} \pi_T \, e^{\,iq\log S_T}g(q) \, \rd q\right), 
\label{interchange}
\end{eqnarray}
from which it follows that
\begin{eqnarray}
\mathbb{E}(\pi_TH_T) = \frac{1}{\sqrt{2\pi}} \, \left( \int_{-\infty}^{\infty} \mathbb{E} \left [\pi_T e^{\,iq\log S_T} \right ] \, g(q)\, \rd q\right).
\label{Derivative price as a fourier portfolio}
\end{eqnarray}
Inserting (\ref{Imaginary Derivative}) into (\ref{Derivative price as a fourier portfolio}), we then have
\begin{eqnarray}
H_0 = \frac{1}{\sqrt{2\pi}} \, \left( \int_{-\infty}^{\infty} F_0(q)\,g(q)\, \rd q\right).
\label{Derivative price as a fourier portfolio final}
\end{eqnarray}
Thus the price of the derivative can be expressed as the value of a portfolio of imaginary power payoff derivatives. 
To check that the interchange of the expectation and the integration in equation 
(\ref{interchange}) is valid  (Kingman \& Taylor 1966, Theorem 6.5), we note that 
\begin{eqnarray}
\mathbb{E} \left(  \int_{-\infty}^{\infty} \pi_T \, e^{\,iq\log S_T}g(q) \, \rd q\right) =
\int_{\omega \in \Omega} \,  \int_{-\infty}^{\infty}   \pi_T \, e^{\,iq\log S_T} g(q) \,  \rd q \, \mathbb{P}(\rd \omega),
\end{eqnarray}
and that
\begin{eqnarray}
\int_{\omega \in \Omega} \, \int_{-\infty}^{\infty}  \left | \pi_T \, e^{\,iq\log S_T} g(q) \right |  \rd q \,\mathbb{P}(\rd \omega)
 = \mathbb E [ \pi_T]  \int_{-\infty}^{\infty}  \left | g(q) \right | \rd q  < \infty.
\end{eqnarray}

As an example we consider a European-style  derivative payoff $H_T = f( \log S_T) $ at time $T$ for which $f$ takes the form of a normal density function
\begin{eqnarray}
f(x) = \frac{1}{\sqrt{2\pi u}}\, e^\mathlarger{{-\frac{1}{2} \, \frac{(x-a)^2}{u}}},
\label{Normal density payoff function}
\end{eqnarray}
with mean $a$ and variance $u$. In the case of a geometric Brownian motion model, the random variable corresponding to the terminal value of the underlying asset is normally distributed with a risk-neutral density of the form
\begin{eqnarray}
\tilde \theta(x) = \frac{1}{\sqrt{2\pi v}}\, e^\mathlarger{{-\frac{1}{2} \, \frac{(x-b)^2}{v}}},
\label{Normal density function GBM}
\end{eqnarray}
where $b = (r - \frac{1}{2}\sigma^2)\,T$ and $v = \sigma^2T$. The price of the derivative at time zero is given by
\begin{eqnarray}
H_0 = e^{- r T}\,\tilde{\mathbb{E}}\,[H_T] = e^{- r T} \int_{-\infty}^{\infty} \tilde \theta(x)  f(x)\, \rd x.
\end{eqnarray}
With some calculation, we find that 
\begin{eqnarray}
H_0 = e^{- r T}\, \frac{1}{\sqrt{2\pi (u+v)}}\, e^{\mathlarger{-\frac{1}{2}\, \frac{(a-b)^2}{u+v}}}.
\label{Normal density function derivative price general formula}
\end{eqnarray}
For instance, if we set $a=0, u=1$ and insert the aforementioned values of $b$ and $v$, we obtain
\begin{eqnarray}
H_0 = e^{- r T}\, \frac{1}{\sqrt{2\pi (1+\sigma^2 T)}}\, e^{\mathlarger{-\frac{1}{2}\, \frac{(r-\frac{1}{2}\sigma^2)^2\,T^2}{1+\sigma^2 T}}}.
\label{Example Normal density function derivative price general formula}
\end{eqnarray}
Alternatively, we can replicate the payoff of the derivative as a portfolio of imaginary power payoffs using the Fourier technique. Since $f \in L^1$, we can set
\begin{eqnarray}
g(q) = \frac{1}{2\pi\sqrt{u}}\, \int_{-\infty}^{\infty} e^{\,-iqx}\, e^\mathlarger{{-\frac{1}{2} \, \frac{(x-a)^2}{u}}} \rd x ,
\end{eqnarray}
and a calculation gives
\begin{eqnarray}
g(q) = \frac{1}{\sqrt{2\pi}}\, e^ {- \frac{1}{2}q^2u}\,e^{-iqa}.
\label{Fourier Transform for Normal density function}
\end{eqnarray}
By (\ref{Derivative fourier payoffs}) and (\ref{Fourier Transform for Normal density function}), and using the fact that $f$ is a good function, one sees that the payoff of the derivative can be expressed in the form
\begin{eqnarray}
H_T = \frac{1}{2\pi} \int_{-\infty}^{\infty}S_T^{iq}\, e^ {- \frac{1}{2}q^2u}\,e^{-iqa}\rd q .
\end{eqnarray}
Then by (\ref{Derivative price as a fourier portfolio final}) we obtain the derivative price as a portfolio of imaginary power-payoff prices:
\begin{eqnarray}
H_0 = \frac{1}{2\pi} \, \left( \int_{-\infty}^{\infty} F_0(q)\, \, e^ {- \frac{1}{2}q^2u}\,e^{-iqa}\, \rd q\right).
\label{Derivative price NDF}
\end{eqnarray}

We observe, finally, that if the prices of imaginary power payoff derivatives delivering the cash flows  defined by (\ref{imaginary power payoff})
are given for all $q \in \mathbb R$, then by adapting the framework of Proposition \ref{determination of Levy exponent} we can work out the implied L\'evy exponent modulo some specified freedom. We recall that the price of a power-payoff derivative at time zero is given for power $q$ by equation  (\ref{value of imaginary power payoff}). Therefore if we consider the function $\{D_0(q)\}_{q \in \mathbb R}$ defined by
\begin{eqnarray}
D_0(q) = \frac {1}{T} \log \frac{F_0 (q)}{S_0^{\, iq}  \, \re^{r(iq-1)T }}, 
\end{eqnarray}
we find that
\begin{eqnarray}
D_0(q) =  -iq \, ( \psi(\sigma -\lambda) -\psi(-\lambda)) + \psi (iq\sigma -\lambda) - \psi(-\lambda ) 
= \tilde{\psi}(iq\sigma) - iq \, \tilde{\psi}(\sigma),
\end{eqnarray}
where $\tilde{\psi}(\alpha) = \psi (\alpha -\lambda) - \psi(-\lambda ) $.
Without of loss of generality we can then set $\sigma = 1$ to get
\begin{eqnarray}
D_0(q) = \tilde{\psi}(iq) - iq \, \tilde{\psi}(1).
\end{eqnarray}
This implies that $\tilde \psi (iq) = D_0(q) + iqb$ for some $b \in \mathbb R$. Now, $\psi (\alpha) = \tilde \psi (\alpha +\lambda) - \tilde \psi (\lambda)$. It follows that for some $\lambda$ and $b$ the L\'evy exponent takes the form
\begin{eqnarray}
\psi (i q) = D_0 (i q +\lambda) - D_0 (\lambda) + ib q.
\end{eqnarray}
Thus, we see that once we have been given a range of price data for imaginary power-payoff options, we can work out the L\'evy exponent modulo a transformation of the form   
\begin{eqnarray}
\psi(i q) \rightarrow \psi(iq  + \mu) - \psi(\mu) + i c q,
\end{eqnarray}
where $c$ and $\mu$ are constants.
\begin{acknowledgments}
\noindent
The authors wish to express their gratitude to D. C. Brody, E. Eberlein, M. Forde, L. Sanchez-Betancourt, U. Schmock, J. Zubelli and an anonymous referee for helpful comments. GB acknowledges support from the George and Victoria Karelia Foundation and from FinalytiQ Limited, Basildon. LPH acknowledges support from the Simons Foundation. This work was carried out in part at the Aspen Center for Physics, which is supported by National Science Foundation grant PHY-1607611. 
\end{acknowledgments}

\end{document}